\documentclass{elsart}

\usepackage{cite}

\begin{document}

\begin{frontmatter}

\title{On the application of homotopy--perturbation and Adomian decomposition
methods to the linear and nonlinear Schr\"{o}dinger
equations}

\author{Francisco M. Fern\'{a}ndez \thanksref{FMF}}

\address{INIFTA (UNLP, CCT La Plata-CONICET), Divisi\'{o}n Qu\'{i}mica Te\'{o}rica,\\
Diag. 113 y 64 (S/N), Sucursal 4, Casilla de Correo 16,\\
1900 La Plata, Argentina}

\thanks[FMF]{e--mail: fernande@quimica.unlp.edu.ar}

\begin{abstract}
I discuss a recent application of homotopy perturbation and Adomian decomposition methods
to the linear and nonlinear Schr\"odinger equations. I propose a generalization of
the procedure for the treatment of a wider class of problems.
\end{abstract}

\end{frontmatter}

\section{Introduction}

The Homotopy Perturbation Method (HPM)\cite
{RDGP07,CHA07,YO07,CH07a,EG07,GAHT07,SNH07,CH08,ZLL08,SNH08b,M08,SNH08,RAH08,YE08,SG08}
(the reader should not mistake this abbreviation with that for the
Hypervirial Perturbation Method\cite{F00}) is supposed to solve any linear
and nonlinear problem in any field of theoretical physics. In a couple of
papers I have shown that some of those results are useless and worthless\cite
{F07,F08}. My criticisms have not been welcomed by the referees (most
probably homotopy devotees) and therefore they remain unpublished.

The purpose of this communication is to analyze a recent application of HPM
and Adomian decomposition method (ADM) to the linear and nonlinear
Schr\"{o}dinger equations\cite{SG08}. Since those equations are known to be relevant
to many fields of physics, any approach to solve them is always welcome.

\section{Linear Schr\"{o}dinger equation}

Sadighi and Ganji\cite{SG08} consider the linear Schr\"{o}dinger equation
\begin{equation}
u_{t}(x,t)+iu_{xx}(x,t)=0  \label{eq:Schro_lin}
\end{equation}
where $i^{2}=-1$ and the subscripts indicate derivatives with respect to the
variables. We immediately recognize this differential equation as a
dimensionless version of the Schr\"{o}dinger equation for the free particle.
Its propagator is well known and there is no reason for any special method
to solve it. However, Sadighi and Ganji\cite{SG08} think otherwise and apply
the HPM to it.

Most textbooks of quantum mechanics solve equation (\ref{eq:Schro_lin}) for
a physical square--integrable initial function $u(x,0)$
\begin{equation}
\int_{-\infty }^{\infty }|u(x,0)|^{2}dx<\infty   \label{eq:normalization}
\end{equation}
however an homotopy devotee does not care about such a small detail. In
fact, Sadighi and Ganji\cite{SG08} cleverly choose some particular
unphysical initial functions, so that they solve the resulting equations
much more easily.

Any undergraduate student in a first course of quantum mechanics knows that
if one chooses an unphysical initial function of the form
\begin{equation}
u(x,0)=\sum_{j}c_{j}e^{\alpha _{j}x}  \label{eq:f(x)_lin}
\end{equation}
then the resulting unphysical solution will be
\begin{equation}
u(x,t)=\sum_{j}c_{j}e^{\alpha _{j}x-i\alpha _{j}^{2}t}  \label{eq:u_lin}
\end{equation}
Notice that the initial function (\ref{eq:f(x)_lin}) does not statisfy (\ref
{eq:normalization}).

The first example chosen by Sadighi and Ganji\cite{SG08}
\begin{equation}
u(x,0)=1+2\cosh (2x)
\end{equation}
is a particular case of (\ref{eq:f(x)_lin}). If you think that it is a
nonsensical quantum--mechanical state then you are not an homotopy devotee.
It follows from equation (\ref{eq:u_lin}) that
\begin{equation}
u(x,t)=1+2\cosh (2x)e^{-4it}  \label{eq:ex1_lin}
\end{equation}
After a masterful application of the HPM Sadighi and Ganji\cite{SG08} obtain
\begin{equation}
u(x,t)=1+2\cosh (2x)\left[ 1-4it+\frac{(4it)^{2}}{2!}+\ldots \right]
\end{equation}
and brilliantly conclude that the limit of this series is equation (\ref
{eq:ex1_lin}).

Example~2 is even simpler than example~1; after some algebraic manipulation
Sadighi and Ganji\cite{SG08} are led to the revelation that when
\begin{equation}
u(x,0)=e^{3ix}
\end{equation}
then
\begin{equation}
u(x,t)=e^{3ix}\left[ 1+9it+\frac{(9it)^{2}}{2!}+\ldots \right] =e^{3(x+3it)}
\label{eq:ex2_lin}
\end{equation}
which is obviously a particular case of equation (\ref{eq:u_lin}). I insist
that if you think that this solution is completely useless from a physical
point of view, then you are not an homotopy devotee.

\section{Nonlinear Schr\"{o}dinger equation}

The nonlinear Schr\"{o}dinger equation
\begin{equation}
iu_{t}+u_{xx}+\gamma |u|^{2}u=0  \label{eq:Sch_nonlin}
\end{equation}
is a much more interesting and challenging problem. However, Sadighi and
Ganji\cite{SG08} cleverly restrict themselves to unphysical solutions with
unit modulus $|u|=1$ so that the problem reduces to a trivial linear
differential equation with constant coefficients:
\begin{equation}
iu_{t}+u_{xx}+\gamma u=0  \label{eq:nonlin->lin}
\end{equation}
If you think that this restriction is ridiculous then I tell you that you
are not an homotopy devotee.

Any student in a first course of calculus will immediately realize that a
particular solution is
\begin{equation}
u(x,t)=e^{i\alpha x}e^{i(\gamma -\alpha ^{2})t}  \label{eq:u_nonlin}
\end{equation}

When $\alpha =1$ and $\gamma =2$ we obtain
\begin{equation}
u(x,t)=e^{i(x+t)}  \label{eq:ex3_u}
\end{equation}
which is exactly the example~3 of Sadighi and Ganji\cite{SG08}. After
masterfully solving the tedious HPM equations Sadighi and Ganji\cite{SG08}
obtain
\begin{equation}
u(x,t)=e^{ix}\left[ 1+it+\frac{(it)^{2}}{2!}+\ldots \right]
\end{equation}
which obviously converges to (\ref{eq:ex3_u}).

When $\alpha =1$ and $\gamma =-2$ equation (\ref{eq:u_nonlin}) reduces to
the solution of example~4:
\begin{equation}
u(x,t)=e^{i(x-3t)}  \label{eq:ex4_u}
\end{equation}
As the reader may have guessed, after solving the HPM equations Sadighi and
Ganji\cite{SG08} obtain the power series
\begin{equation}
u(x,t)=e^{ix}\left[ 1-3it+\frac{(3it)^{2}}{2!}+\ldots \right]
\end{equation}
that clearly converges towards (\ref{eq:ex4_u}).

It is pointless telling an homotopy devotee that all those calculations are
useless and worthless, he is content with showing that his calculation
yields the exact result known to everybody.

The reader should not think that Sadighi and Ganji\cite{SG08} merely
restrict to the calculation of the power series of exponential functions $%
e^{ibt}$ by means of HPM. They also do exactly the same by means of another
of the most fashionable approaches: the Adomian decomposition method
(ADM)\cite{SG08}.

\section{Conclusions}

The discussion of the preceding sections show that Sadighi and Ganji\cite
{SG08} study the simplest linear and nonlinear Schr\"{o}dinger equations.
They choose unphysical solutions that do not correspond to actual
quantum--mechanical states but are more tractable. One such choice converts
the nonlinear Schr\"{o}dinger equation into a linear one and consequently
the authors never solve the nonlinear problem. Then they apply two
approximate methods, the HPM and ADM, and obtain a power series
approximation to the solutions. As shown above, the calculation reduces to
the Taylor expansion of exponential functions of the form $e^{ibt}$ about $%
t=0$. If the reader thinks that it is not a great achievement I will prove
him/her wrong. First of all, notice that the authors do not use the obvious
formula
\begin{equation}
u(x,t)=\sum_{j=0}\frac{t^{j}}{j!}\left. \frac{\partial ^{j}u}{\partial t^{j}}%
\right| _{t=0}
\end{equation}
but apply two cumbersome methods in order to obtain this expansion. Second,
they managed to publish their remarkably useless and nonsensical
contribution in a research journal\cite{SG08}. In another contribution I
will analyze how one can do it.

Finally, I will like to propose a generalization of the great achievement of
Sadighi and Ganji\cite{SG08}. Instead of the simple linear equation (\ref
{eq:Schro_lin}) one may consider the more general one
\begin{equation}
u_{t}+i\hat{A}u=0  \label{eq:lin_gen}
\end{equation}
where $\hat{A}$ is a linear operator. According to the textbooks on quantum
mechanics, its formal solution is
\begin{equation}
u=e^{-it\hat{A}}u_{0}  \label{eq:lin_gen_u}
\end{equation}
where $u_{0}=u(t=0)$. A successful application of the HPM or ADM will lead
to the series
\begin{equation}
u=\left[ 1-it\hat{A}+\frac{(-it\hat{A})^{2}}{2!}+\ldots \right] u_{0}
\label{eq:lin_gen_u_series}
\end{equation}

If you do not succeed in publishing your results you may try a simpler
problem; for example, when
\begin{equation}
\hat{A}u_{0}=au_{0}
\end{equation}
you have
\begin{equation}
u=e^{-ita}u_{0}
\end{equation}
and
\begin{equation}
u=\left[ 1-ita+\frac{(-ita)^{2}}{2!}+\ldots \right] u_{0}
\end{equation}
Hopefully, the referee (most probably an homotopy devotee) will understand
your simplified problem and then accept your manuscript.

If your manuscript is accepted then you become a member of the homotopy club
and then you can try something bolder, for example
\begin{equation}
u_{0}=\sum_{k}c_{k}f_{k}
\end{equation}
where
\begin{equation}
\hat{A}f_{k}=a_{k}f_{k}
\end{equation}
I leave the result as an exercise for the future homotopy devotee.

Notice that you can thus write a paper for every solvable or unsolvable
problem which you can think of. Most probably, all of them will be accepted
in any one of the homotopy journals\cite{RDGP07,CHA07,YO07,CH07a,EG07,GAHT07,
SNH07,CH08,ZLL08,SNH08b,M08,SNH08,RAH08,YE08,SG08}
(unless you dare to criticize another
homotopy result).

\end{document}